\documentclass[%
 reprint,
superscriptaddress,
 amsmath,amssymb,
 aps,
]{revtex4-1}

\usepackage[utf8]{inputenc}
\usepackage[english]{babel}
\usepackage[usenames, dvipsnames]{color}
\usepackage{graphicx}
\usepackage{dcolumn}
\usepackage{bm}

\usepackage{amsmath}	
\usepackage{amssymb}	

\def\be{\begin{equation}}
\def\ee{\end{equation}}

\newcommand{\ptrans}{p_{\rm trans}}
\newcommand{\ntrans}{\rho_{\rm trans}}

\newcommand{\etrans}{\varepsilon_{\rm trans}}

\newcommand{\De}{\Delta}
\newcommand{\ep}{\varepsilon}

\newcommand{\cQMsq}{c^2_{\rm QM}}

\begin{document}

\title{Can magnetic fields stabilize or destabilize twin stars?}
\date{\today}
\author{R.O. Gomes}
\affiliation{Frankfurt Institute for Advanced Studies,
Frankfurt am Main, Germany}
\email{gomes@fias.uni-frankfurt.de}
\author{V. Dexheimer}
\affiliation{Department of Physics, Kent State University, Kent OH 44242 USA}
\author{S. Han}
\affiliation{Department of Physics and Astronomy, University of Tennessee, Knoxville  TN 37996, USA}
\affiliation{Department of Physics and Astronomy, Ohio University, Athens OH~45701, USA
}
\author{S. Schramm}
\affiliation{Frankfurt Institute for Advanced Studies,
Frankfurt am Main, Germany}

\begin{abstract}
Sharp phase transitions described by stiff equations of state allow for the existence of a third family of stable compact stars (besides white dwarfs and neutron stars), twin stars. 
In this work, we investigate for the first time the role of strong magnetic fields on non-magnetic 
twin stars sequences and the case in which magnetic fields themselves give rise to a third family of stable stars. 
We use three sets of equations of state to study such effects from a more general point of view: the Quark-Hadron Chiral Parity-Doublet (Q$\chi$P) model for both hadronic and quark phases, and 
the Many-Body Forces (MBF) model connected to either the MIT Bag model with vector interaction (MIT) or to the Constant-Sound-Speed (CSS) approximation for the quark phase, through a Maxwell construction. 
Magnetic field effects are introduced in the structure of stars through the solution of the Einstein-Maxwell equations, assuming a poloidal magnetic field configuration and a metric that allows for the description of deformed stars.
We show that strong magnetic fields can destabilize twin star sequences, with the threshold intensity being model dependent. On the other hand, magnetic fields can also give rise to twin stars in models that did not predict these sequences, up to some point when they are again destabilized. In this sense, magnetic fields can play an important role on the evolution of neutron stars.

\end{abstract}

\maketitle

\section{Introduction}

The idea of a third family of compact stars with small radii (tertiary stars) was first suggested by Gerlach in 1968 \cite{Gerlach:1968zz} in a generic context, and then by K\"ampfer in 1981 in the context of hybrid stars with a quark core \cite{Kampfer:1981yr}. In the past years, the interest in these stars increased due to studies indicating that neutron star radii might be smaller than previously expected. 
Recently, this idea came back in order to explain data from neutron star mergers suggesting again smaller radii \cite{TheLIGOScientific:2017qsa}. 

In addition, the observation of twin stars (two stars with the same mass and significantly different radius) 
would be a definite confirmation of a strong first-order phase transition in stars, as already pointed out in Refs. \cite{Glendenning:1998ag,Schertler:2000xq,Alvarez-Castillo:2013cxa,Benic:2014jia}. 
This is a very timely idea, as NASA’s Neutron star Interior Composition Explorer (NICER) has been attached to the space station in June of 2017 and will soon report accurate data of neutron star radii. 
The third family of stars has also been investigated with regard to different properties and contexts such as supernovae \cite{Hempel:2015vlg,Heinimann:2016zbx}, pasta phases \cite{Alvarez-Castillo:2014dva}, particle populations \cite{Banik:2001yw,Chatterjee:2009xi,Pagliara:2014gja}, color superconductivity \cite{Banik:2002kc,Alford:2017qgh}, rotation \cite{Banik:2004ju,Bejger:2016emu}, and tidal deformation in neutron star mergers \cite{Paschalidis:2017qmb,Alvarez-Castillo:2018pve,Gomes:2018eiv}.


The conversion mechanism of hadronic stars into hybrid or quark stars is still an open question, and the possibility of a third family or even two families of stars have been addressed in past works \cite{Glendenning:1998ag,Drago:2015dea,Drago:2015cea,Alvarez-Castillo:2016dyz}. 
From the microscopic point of view, a soft equation of state (EoS) in the transition region is necessary. It  generates a large energy density gap which creates a sequence of hybrid stars that is disconnected from the hadronic branch (third family) \cite{Benic:2014jia}.
On the other hand, in order to reproduce the observational stellar mass constraints, it is necessary that both hadronic and quark matter equations of state are stiff. For the EoS criteria for hybrid star branches to be connected or disconnected to hadronic ones, see  Refs. \cite{Alford:2013aca,Alford:2015gna,Alford:2015dpa}.

Moreover, it is important to note that magnetic fields are also relevant in the calculation of the structure of neutron stars. It has been shown that although they do not significantly affect  the matter equation of state, strong magnetic fields can change the macroscopic properties of neutron stars dramatically \cite{Franzon:2015sya,Gomes:2017zkc,Chatterjee:2018prm}. 
In particular, they have the effect of decreasing the central density of stars due to Lorentz force, turning hybrid and hyperonic stars into nucleonic stars (composed only of nucleons and leptons) \cite{Franzon:2015sya,Franzon:2016urz}. 

In this work, we address for the first time how strong magnetic fields affect the mass-radius relation of neutron stars, both generating and destabilizing twin-star configurations. 
We start from describing different models that generate twin configurations without resorting to magnetic fields. Then, we discuss the effect of adding strong magnetic field effects to these twin stars. After that, we discuss the possibility of twin stars that exist only in the case when strong magnetic fields are present. Finally, we present our conclusions and outlook.

\begin{table*}
  \caption{\label{Hmodels} Nuclear matter properties at saturation for the hadronic models used in this analysis, assuming a saturation density of $\rho_0=0.15\,\rm{fm}^{-3}$. The columns read: model, nucleon effective mass $m^*_n$, compressibility modulus $K_0$, binding energy per nucleon $B/A$,  and symmetry energy $a_{sym}$,  at nuclear saturation density $\rho_0$.}

\begin{center}
\begin{tabular}{ccccc}
 \hline
Model & $\quad m^*_n/m_n$ &  $\quad K_0$ & $\quad B/A$ & $a_{sym}$    \\
 &  &  $\quad (\mathrm{MeV})$ & $\quad (\mathrm{MeV})$ & $\quad (\mathrm{MeV})$     \\

\hline \hline

1. MBF \cite{Gomes:2014aka} & 0.66 & 297  & -15.75  & 32.00   \tabularnewline
2. Chiral \cite{Dexheimer:2014pea} & 0.67 & 318.76  & -15.65  & 32.43  \tabularnewline

\hline\hline
  \end{tabular}
\end{center}
\end{table*}

\begin{figure}[!ht]  
  \includegraphics[width=.9\linewidth]{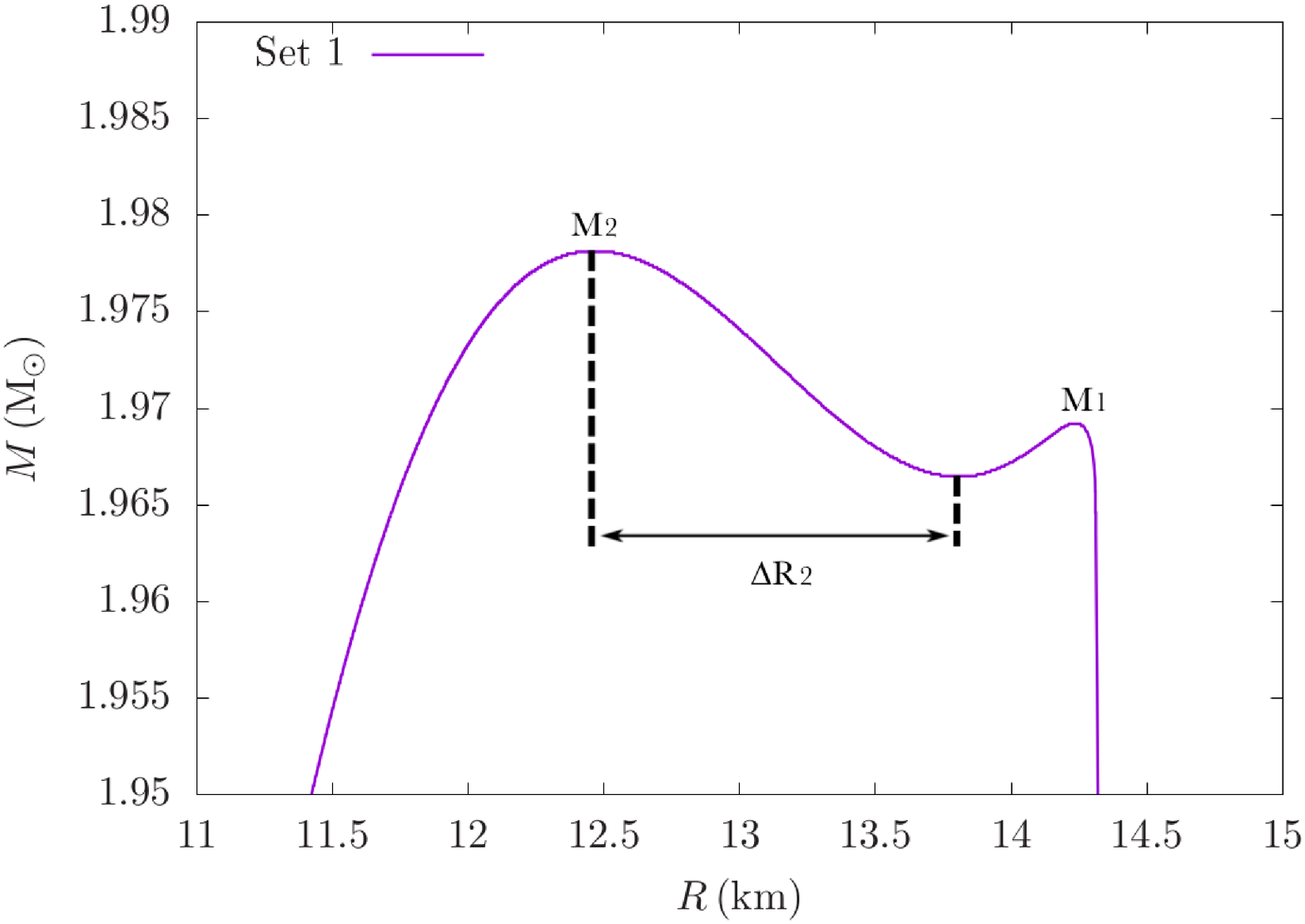}
  \includegraphics[width=.9\linewidth]{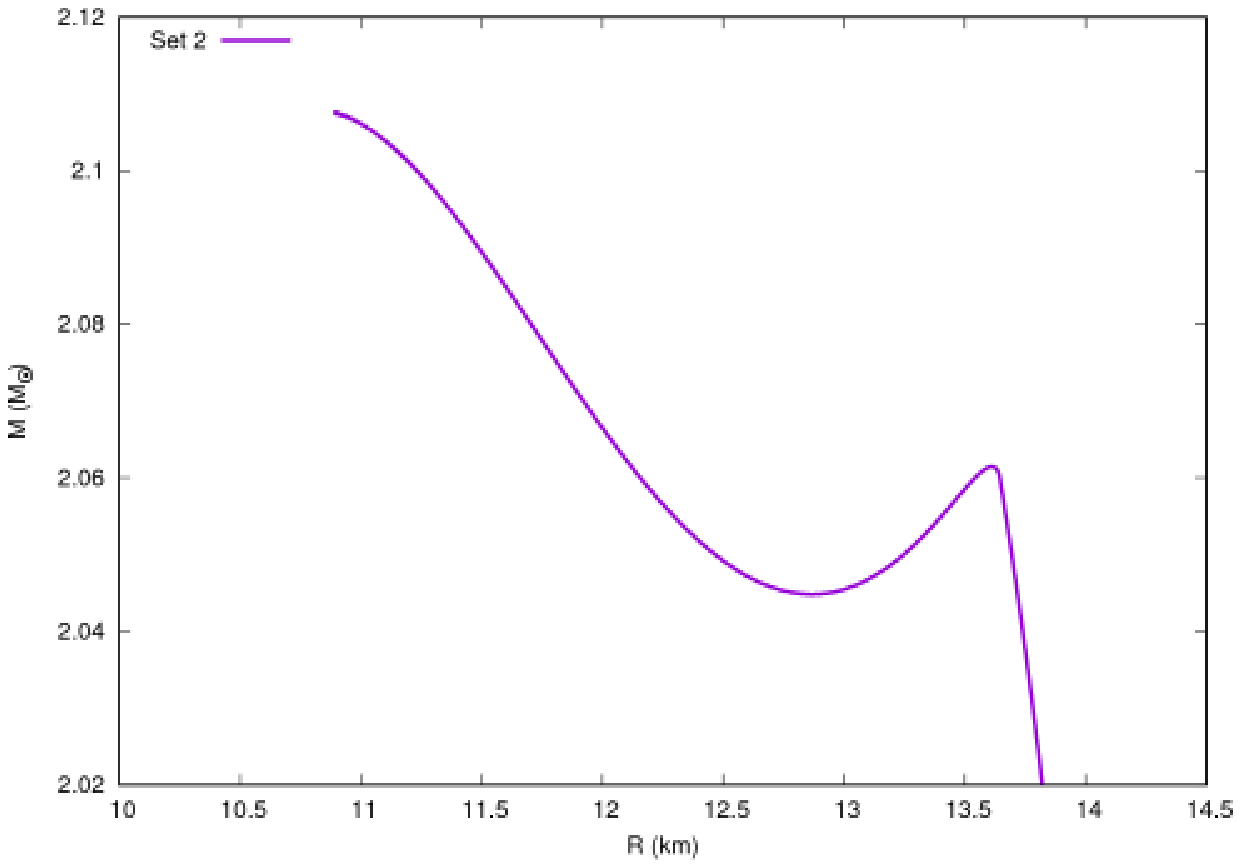}
  \includegraphics[width=.9\linewidth]{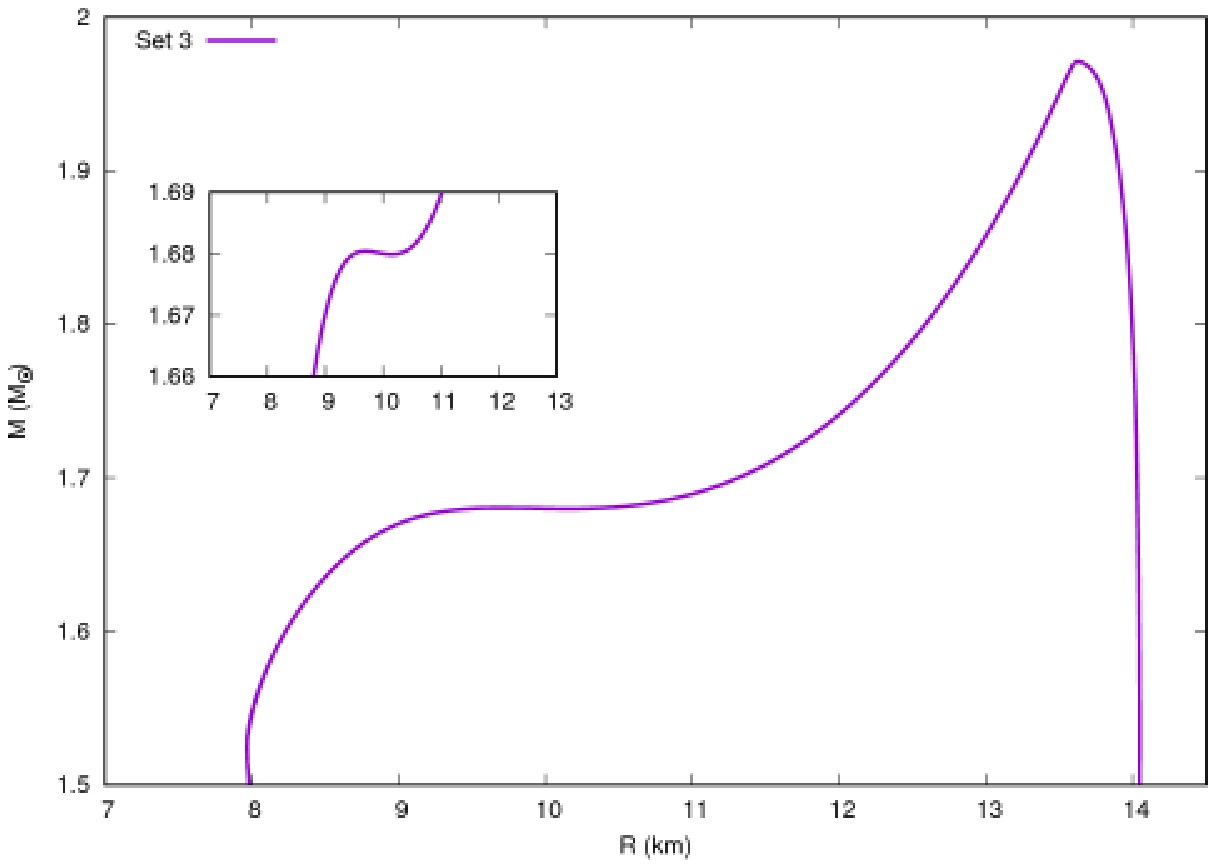}
  \caption{Mass-radius diagram for Sets 1, 2, and 3, all 
reproducing twin-star configurations (without the inclusion of magnetic fields). $M_1$ and $M_2$ are the hadronic and hybrid branch maximum masses, and $\Delta R_2$ is the radius interval covered by the third family.}
\label{nonB-twins}
\end{figure}

\section{Formalism}

In this section, we present the formalism used to study the effects of magnetic fields in twin stars. In subsection \ref{non-mag_twins}, we present two equations of state that allow for the existence of a third family of non-magnetic stars. Assuming that a first-order phase transition takes place at high densities, matter is modeled in two different scenarios: one caused by a deconfinement phase transition to quark matter, and another one caused by a phase transition to high-strangeness matter. We use two relativistic mean field models to describe hadronic matter, considering a nuclear saturation density $\rho_0=0.15\,\rm{fm}^{-3}$,  parametrized to reproduce the nuclear properties at saturation shown in Table \ref{Hmodels}. All models/constructions we use are able to reproduce standard nuclear and astrophysical constraints \cite{Alford:2013aca,Gomes:2014aka,Franzon:2016urz,Gomes:2018eiv}. 
In subsection II B, we investigate the effects of strong magnetic fields on the twin stars from section II A. In section II C, we present a configuration of highly magnetized hybrid stars which only in this case generates magnetic twin-star sequences. For the latter analysis, only the hadron-quark phase transition scenario is considered, using one of the models from section I A but with a different parametrization.

\begin{table*}
  \caption{\label{TwinProp} Properties of non-magnetic twin stars for the sets used in this analysis. The columns are: masses and radii of the maximum mass star at the first (hadronic) and second (hybrid) branches, $M_1$ and $R_1$, and $M_2$ and $R_2$, respectively; and lastly the range of radii $\Delta R_2$ of stable third-family stars. These quantities are displayed in the top panel of Figure 1.}

\begin{center}
\begin{tabular}{cccccc}
 \hline
Set & $\quad M_1$ &  $\quad R_1$ & $\quad M_2 $ & $\quad R_2$ & $\quad \Delta R_2$   \\
 &  $\quad(\mathrm{M_{\odot}})$ & $\quad (\mathrm{km})$ & $\quad (\mathrm{M_{\odot}})$ & $\quad (\mathrm{km})$  & $\quad (\mathrm{km})$    \\

\hline \hline

Set 1  &$\quad 1.969$ & $\quad 14.23$  & $\quad 1.978$  &$\quad 12.47 $& $\quad  1.33$ \tabularnewline
Set 2  & $\quad 2.061$ & $\quad 13.61$  & $\quad 2.108$  & $\quad 10.9$  &  $\quad 1.97 $ \tabularnewline
Set 3  & $\quad 1.971 $ & $\quad 13.63$  & $\quad 1.680$  & $\quad 9.70$  &  $\quad 0.45 $ \tabularnewline
\\
\hline\hline
  \end{tabular}
\end{center}
\end{table*}

\subsection{Non-magnetic twin stars}\label{non-mag_twins}

In the following, we present different relativistic mean field models that are used to generate equations of state taking into account charge neutrality and chemical equilibrium. Sequences of spherical, non-rotating and non-magnetic twin stars are then calculated by solving the Tolman-Oppenheimer-Volkoff equations. 

\subsubsection{Due to deconfinement}

First, we discuss the case in which a non-magnetic
third family is generated by a hadron-quark phase transition. For such a case, we use an EoS parametrization of the many-body forces (MBF) model for the hadronic phase \cite{Gomes:2014aka} and the MIT bag model with vector interaction for the quark phase (see,  for example Ref. \cite{Franzon:2016urz}). 

The hadronic phase is assumed to consist of only nucleons and leptons. In this relativistic mean-field framework, many-body forces contributions are introduced in the baryon couplings to the scalar fields ($\sigma,\delta$), and are controlled by a parameter $\zeta$. The vector interaction introduced in the MIT bag model is equivalent to the approaches proposed in Refs. \cite{Alford:2004pf,Weissenborn:2011qu,Klahn:2015mfa}, in which is referred to as vMIT or vBag-model, and allows for a stiff EoS for quark matter, able to describe massive hybrid stars. The two phases are connected by a Maxwell construction, which describes a necessary sharp phase transition to quark matter.

In this work, we choose the many-body forces parameter to be $\zeta=0.040$, which is the stiffest possible realistic parametrization of the model (see Table \ref{Hmodels}). The values of the vector coupling, bag constant and mass of the strange quark in the MIT bag model that give rise to a third family are $(g_V/m_V)^ 2=1.7\, \mathrm{fm^ 2}$, $B^{1/4}=171\,\mathrm{MeV}$ and $m_s=150\,\mathrm{MeV}$, respectively. Note that increasing the strange quark mass or repulsion does  not favor hybrid stars, as higher transition pressure and larger energy gap, make it more likely that the stars will become unstable \cite{Ranea-Sandoval:2015ldr}. For Set 1, we have used the BPS equation of state  \cite{Baym:1971pw} for the crust.

For Set 2, we again describe hadronic matter with the same parametrization of the MBF model, but also allowing for hyperon degrees of freedom to appear, reproducing the respective values for the hyperon potentials \cite{Gomes:2014aka}: $U_\Lambda=-28$ MeV, $U_\Sigma=30$ MeV, and $U_\Xi=-18$ MeV. For quark matter, we use the ``constant-sound-speed (CSS)'' parametrization which assumes that the speed of sound in quark matter is pressure-independent for pressures ranging from the first-order transition pressure up to the maximum central pressure of a neutron star \cite{Zdunik:1987,Alford:2013aca}. For a given hadronic matter EoS, CSS parameters are then the pressure at the transition $\ptrans$ (or equivalently the transition density $\ntrans$), the
discontinuity in energy density at the transition $\De\ep$, and the
speed of sound in the high-density phase $\cQMsq$. The CSS parameter values we applied for  the MBF model are $\ntrans=3.5\,\rho_0$, $\De\ep/\etrans=0.3$ and $\cQMsq=1$, where $\etrans\equiv\ep_{\rm MBF}(\ptrans)$. The properties of the third family configuration found for this set is also shown in Table \ref{TwinProp}.
For the crust in Set 2, we have used EoS' from \citet{Baym:1971pw} and
\citet{Negele73ns}.

The corresponding sequence of twin stars for these models are shown in Figure \ref{nonB-twins} and are labeled Set 1 and Set 2.
Increasing the central density, the hadronic stars become more compact due to the larger gravitational attraction.
When a sharp phase transition to quark matter takes place, stars become unstable until the quark core (described with a stiff quark matter EoS) becomes large enough to overcome the instability, creating the third family of stars.
We define the maximum mass for the hadronic and hybrid branches as $M_1$ and $M_2$, respectively. The radius interval for the third family branch, from the minimum between the mass peaks until the hybrid maximum mass peak ($M_2$) is defined as $\Delta R_2$. We display the main properties of these sets in Table \ref{TwinProp}.

\subsubsection{Due to strangeness}

Here, we discuss the case in which there are stable twin stars (not considering magnetic fields) still generated by a strong phase transition not necessarily related to deconfinement. The corresponding EoS is modeled using the Quark-Hadron Chiral Parity-Doublet (Q$\chi$P) model \cite{Steinheimer:2011ea,Dexheimer:2012eu}, which contains positive and negative parity states of the baryons octet as well as quarks within the mean field approximation. The introduction of an excluded volume for the baryons suppresses hadrons at high density and/or temperature, allowing the quarks to dominate \cite{Steinheimer:2010ib}.
The coupling constants of the model were fitted to reproduce the vacuum masses of the baryons and mesons, and reasonable values for the hyperon potentials ($U_\Lambda=-30.44$ MeV, $U_\Sigma=2.47$ MeV, and $U_\Xi=-26.28$ MeV). In addition to the properties presented in Table \ref{Hmodels}, the vacuum expectation values of the scalar mesons are constrained by reproducing the pion and kaon decay constants $f_\pi$ and $f_\kappa$. The BPS equation of state \cite{Baym:1971pw} is also used to describe the crust of these stars.

The effects of strangeness on the quark sector of this equation of state are studied in Ref.~\cite{Dexheimer:2014pea} by varying the quark coupling to the strange vector meson. 
In particular, it is found that a large amount of strange quarks is related to a softer equation of state that posses a first-order phase transition observed in a reduction of the chiral condensate. Note that in this model both phases contain hadrons (nucleons and hyperons) and quarks and there would not be a first-order phase transition (but a crossover instead) if it were not for the choice of strange quark coupling. In this way, a phase transition separates a phase with lower strangeness from a phase with larger strangeness.  
In the context of hadronic matter, the possibility of smooth and strong phase transitions to strange matter has been explored in Ref.~\cite{Gulminelli:2013qma}. 
We label this sequence of twin stars Set 3 (see Figure \ref{nonB-twins}).
For example, a particular star mass of $1.68\,\mathrm{M_\odot}$, corresponding to radii 
of  $14.00$ km and $9.60$ km in different branches, contains strangeness fraction of $f_s = 0.01$ and $f_s = 1.68$, respectively, at the center of the star in each branch. The strangeness is defined as $f_s = \sum_i \rho_i Q_{S_i} / \rho_B$, where $\rho_i$ is the number density of each particle, $\rho_B$ the baryon number density, and $Q_{S_i}$ is the strangeness of each particle. The properties of the two branches of stars are also shown in Table \ref{TwinProp}.

\subsection{Adding magnetic field effects}\label{mag_twins}

In this section, we investigate the effects of magnetic fields on the twin stars presented in section II A. 
Magnetic field effects can be introduced simultaneously in the macroscopic structure of neutron stars by the solution of the Einstein-Maxwell equations and in the microscopic formalism of the EoS through Landau quantization. Nevertheless, as already shown in Refs. \cite{Chatterjee:2014qsa,Gomes:2017zkc}, the latter does not show significant 
effects on the macroscopic properties of stars for magnetic fields of $\sim 10^{18}\,\mathrm{G}$ or smaller (which is the case in this work) and, therefore, is not taken into account in this work.

The macroscopic stellar structure, on the other hand, is significantly modified by magnetic fields with strengths $\sim 10^{18}\,\mathrm{G}$ in the stellar center  \cite{Bonazzola:1993zz}. For this reason, spherical solutions of Einstein's equations must be abandoned when studying strongly magnetized stars. For this purpose, we make use of the LORENE C++ class library for numerical relativity that generates equilibrium configurations from Einstein-Maxwell’s field equations assuming a poloidal magnetic field configuration produced self-consistently by a macroscopic current \cite{Bonazzola:1993zz}. The magnetic field strengh depends on the stellar radius (with respect to the symmetry axis), the dipole magnetic moment and the EoS. For a fixed dipole magnetic moment, the magnetic field increases slowly in the polar direction  towards the center of each star \cite{Dexheimer:2016yqu,Chatterjee:2018prm}.

\begin{figure}[!ht]  
  \includegraphics[width=.9\linewidth]{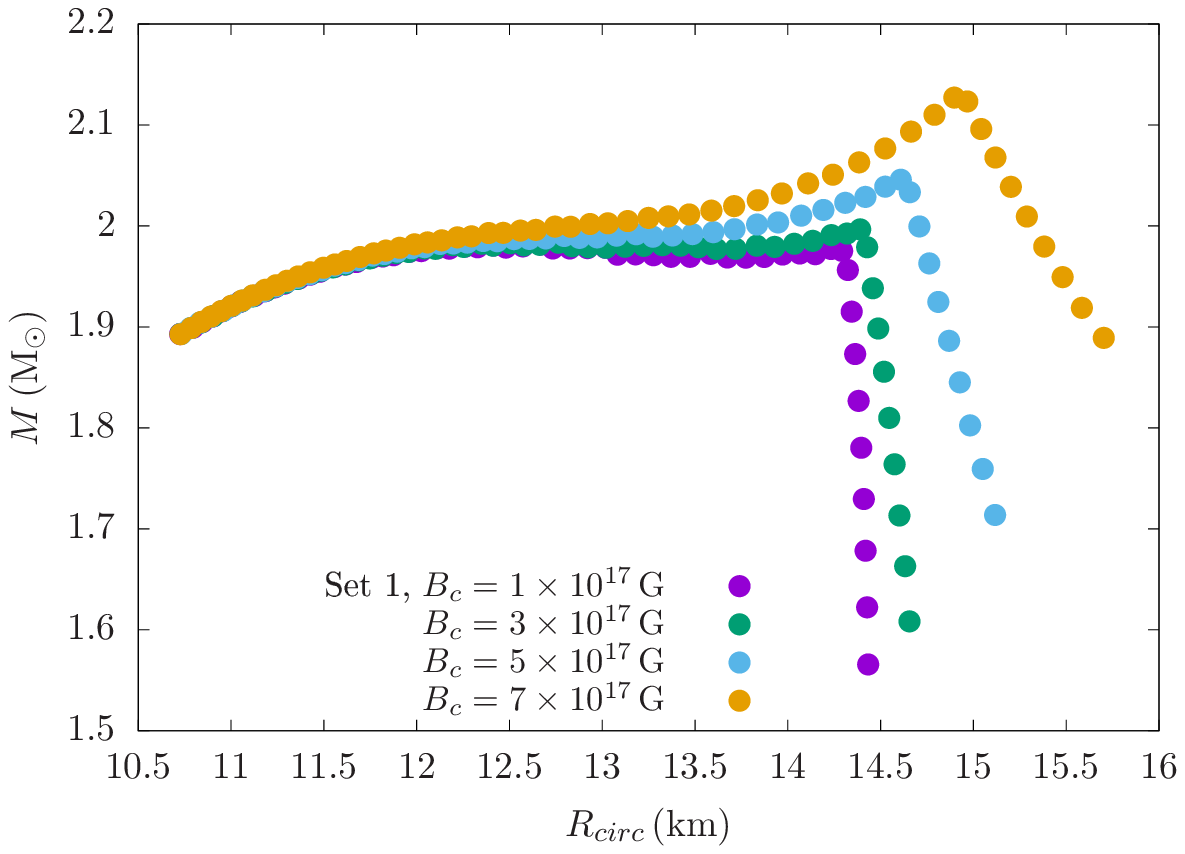}
  \includegraphics[width=.9\linewidth]{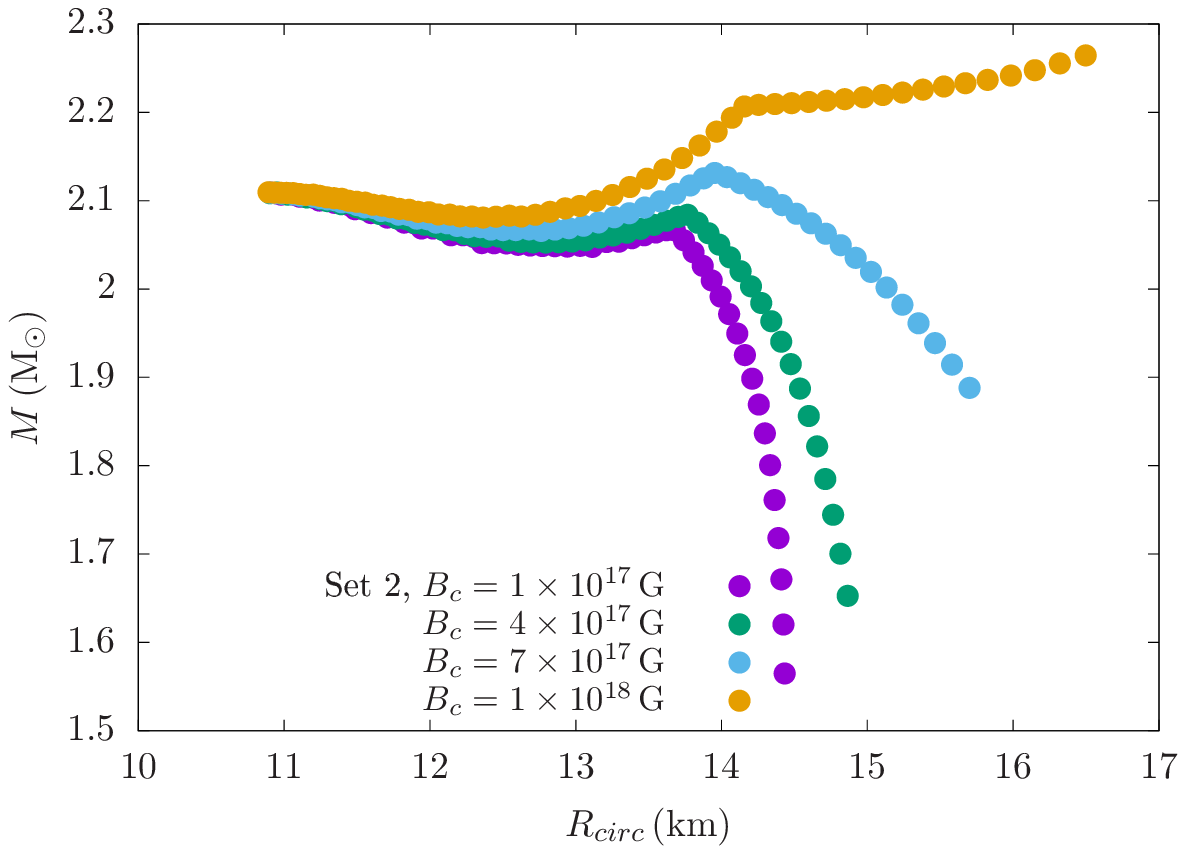}
  \includegraphics[width=.9\linewidth]{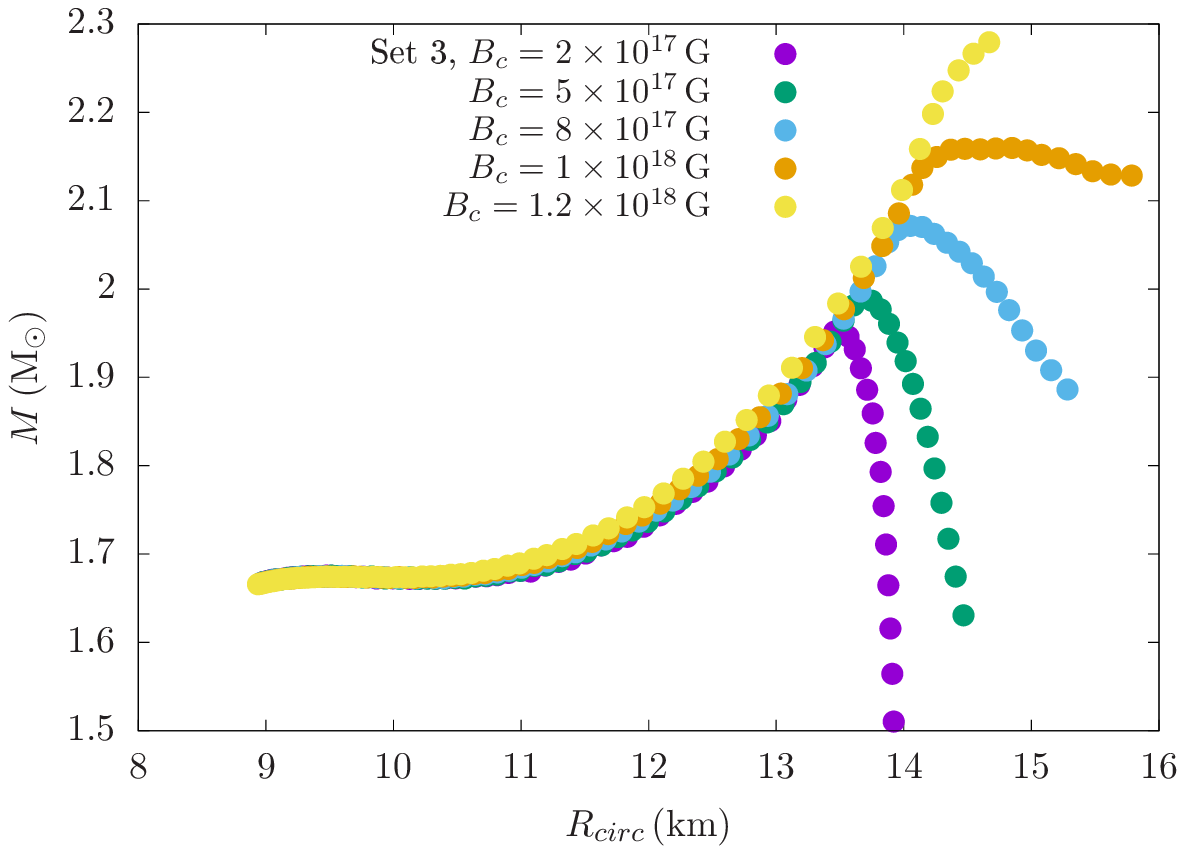}
  \caption{Mass-radius diagram for Sets 1 (top), 2 (middle), and 3 (bottom), for different magnetic field configurations. For Set 1, twin stars disappear for a central magnetic field of around  $B_c\sim 5\times 10^{17}\,\mathrm{G}$, but do not disappear for the maximum central magnetic field configurations investigated in the analysis of Sets 2 and 3.}
\label{magtwins_sets}
\end{figure}

In this work, for the first time, we fix the magnetic field in the center of each star of the sequence by adjusting the dipole magnetic moment for each stellar central density. This is shown in Figure \ref{magtwins_sets} for Set 1-3 of section II A. 
Note that in this work we have chosen to discuss twin stars from the analysis of their (magnetic-axis) equatorial radii.  This is because neutron star radius measurements usually refer necessary to their equatorial radii. In the case of measurements using x-ray bursts, as for example the ones to be performed by NICER, they refer to stellar equatorial radius (possibly at non-zero latitudes) defined by rotation \cite{Ozel:2015ykl}. In the case of measurements from gravitational waves, they refer to a radius defined by the projection from the binary orbital motion, since this is the direction in which deformation takes place \cite{Bauswein:2017vtn,Raithel:2018ncd}. In any case, all these “equators” are not expected to be significantly different from each other angular wise in comparison to the polar angle, as magnetic and rotation axes are expected to be almost aligned for stars with strong magnetic fields \cite{Lander:2018und}.

For Set 1, shown at the top panel in Figure \ref{magtwins_sets}, the introduction of stellar magnetic fields at first increases the masses and radii of stars until after it reaches a central value of $B_c\sim 5\times 10^{17}\,\mathrm{G}$, when it completely destabilizes all twin stars. The change in mass and radius is larger for hadronic stars than for hybrid ones, as hybrid stars are more compact and, therefore, less deformable by magnetic fields \cite{Gomes:2017zkc}. For the hadronic maximum mass star,  $R_{1}=14.23\,\mathrm{km}$ goes to $R_{1}=14.49\,\mathrm{km}$ and  $M_{1}=1.969\,\mathrm{M_{\odot}}$ goes to $M_{1}=2.016\,\mathrm{M_{\odot}}$, for central magnetic fields going from zero to $B_c\sim 4\times 10^{17}\,\mathrm{G}$. 

The behavior of twin stars for Set 2 is not distinct from Set 1, and it is shown at the middle panel in Figure \ref{magtwins_sets}. The main difference is that now none of the physical magnetic fields studied destabilizes the twin stars. This is due to the fact that the original mass range of the twins is larger than for Set 1. The changes in maximum masses and radii for the maximum mass hadronic star, from the non-magnetic case to a central magnetic field $B_c\sim 1\times 10^{18}\,\mathrm{G}$ are: from $R_{1}=13.61\,\mathrm{km}$ to $R_{1}=14.173\,\mathrm{km}$ and from $M_{1}=2.061\,\mathrm{M_{\odot}}$ to $M_{1}=2.208\,\mathrm{M_{\odot}}$.
 
The twin stars from Set 3 are shown at the bottom panel in Figure \ref{magtwins_sets}. In this case, again, none of the physical magnetic fields used in this analysis destabilizes the twin stars, as they are extremely compact. For this set, the changes in maximum masses and radii due to magnetic fields of $B_c\sim 1\times 10^{18}\,\mathrm{G}$ for the maximum mass hadronic star  are from $R_{1}=13.63\,\mathrm{km}$ (non-magnetic) to  $R_{1}=14.65\,\mathrm{km}$ and from $M_{1}=1.971\,\mathrm{M_{\odot}}$ (non-magnetic) to $M_{1}=2.159\,\mathrm{M_{\odot}}$. 

It is important to notice that for central magnetic fields higher than $B_c\sim 1\times 10^{18}\,\mathrm{G}$, the hadronic star branch becomes unstable even before the twin stars for Sets 2 and 3. In these cases, the third family still remains, but the intensity of magnetic fields reaches both numeric and physical limit for our description of magnetic neutron stars. Numerically, beyond this (model dependent) threshold, the purely magnetic contribution to the energy density exceeds the matter part, and the code stops converging. From the physical point of view, stars on the hadronic branch have low central densities, and magnetic field effects on the crust equation of state should be taken into account. The latter is beyond the scope of this work.




\subsection{Magnetic twin stars}\label{mag_twins}

\begin{figure}  
  \centering
  \includegraphics[width=.9\linewidth]{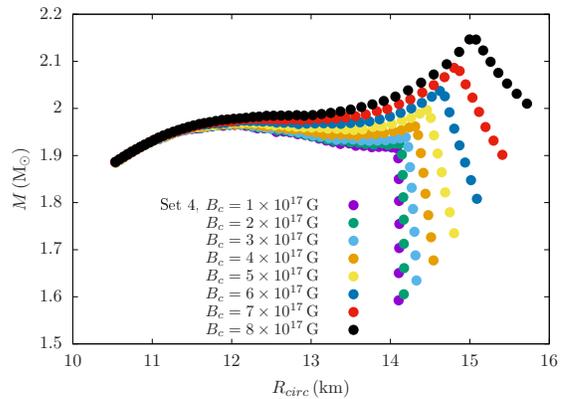}
  \caption{Mass-Radius diagram for Set 4. This configuration does not present twins for the non-magnetic case, and the appearance of twin stars takes place for magnetic fields as high as $B_c\sim 4\times 10^{17}\,\mathrm{G}$. For even higher magnitudes of central magnetic fields, of around $B_c\sim 7\times 10^{17}\,\mathrm{G}$, the third family becomes unstable. }
\label{magtwins_rosana_hybrid}
\end{figure}

In this section, we investigate the effects of magnetic fields generating twin-star configurations that otherwise would not exist. For this purpose, we once more describe a hadron-quark phase transition with a Maxwell construction for the combination of the MBF and MIT models.
The hadronic phase parameterization is the same as in  Section II A 1 ($\zeta=0.040$), but we take the values of the vector coupling and bag constant to be: $(g_V/m_V)^ 2=2.2\,\mathrm{fm^ 2}$ and $B^ {1/4}=160\,\mathrm{MeV}$, in order to reproduce massive and stable hybrid stars \emph{without} the existence of twins (Set 4). The results are presented in Figure \ref{magtwins_rosana_hybrid}.

As the central magnetic field in the hybrid star increases (and so does the magnetic field strength throughout the star), a second mass peak appears, and hence a twin-star configuration. The corresponding threshold is $B_c=3\times 10^ {17}\,\mathrm{G}$ for this configuration. For larger magnetic fields, the masses and radii of stars (especially hadronic ones) keep increasing until $B_c\sim 7\times 10^ {17}\,\mathrm{G}$, when the twin-star branch disappears again, equivalently to the discussion for Set 1 in the last session.

More specifically, the non-magnetic configuration for Set 4 has a maximum mass hybrid star of $M_{2}=1.966\,\mathrm{M_{\odot}}$ and radius $R_{2}=12.00\,\mathrm{km}$. The critical mass star, beyond which all stars are hybrid, has a mass and radius of $M_{c}=1.908\,\mathrm{M_{\odot}}$ and radius $R_{c}=14.05\,\mathrm{km}$.
When magnetic fields are introduced, due to the appearance of a third family of stars at $B_c=3\times 10^ {17}\,\mathrm{G}$, the critical mass star $M_c$ becomes the maximum mass star at the hadronic branch $M_1$, having mass and radius of $R_{1}=14.19\,\mathrm{km}$ and $M_{1}=1.939\,\mathrm{M_{\odot}}$, for this central magnetic field. The interval of radii for the third family $ R_2$ ranges from $13.73\,\mathrm{km}$ to $12.04\,\mathrm{km}$ ($\De R_2=1.69\, \rm{km}$).

Once the third family is established, $\Delta R_2$ decreases as a function of the central magnetic field and ultimately the twin stars disappear. In particular, at $B_c=6\times 10^ {17}\,\mathrm{G}$, when the twin-star configuration is close to the vanishing threshold, the radius and mass of the hadronic maximum mass star is 
$R_{1}=14.62\,\mathrm{km}$ and $M_{1}=2.037\,\mathrm{M_{\odot}}$, and the radius interval for the third family is reduced to $\Delta R_2=(13.32-12.19)\, \rm{km}$.

Note that throughout this work, only gravitational masses were displayed in all figures. Nevertheless, baryonic mass plots would show the same qualitative behavior for twin stars.

\section{Discussion and Conclusions}

In this work we made use of different hadronic and quark models to study the effects of magnetic fields on twin stars generated both without and exclusively by magnetic field effects. The hadronic models employed were the many-body forces (MBF) and the Quark-Hadron Chiral Parity Doublet (Q$\chi$P) model. The quark models were the vector MIT bag model, the Constant-Sound-Speed (CSS) approximation and, once more, the Quark-Hadron Chiral Parity Doublet (Q$\chi$P) model. 


By defining $M_1$ and $M_2$ as the maximum mass stars on the hadronic and hybrid branches, respectively, we have shown that strong magnetic fields do not affect $M_2$ substantially. On the other hand, $M_1$ corresponds to stars that are less compact and, therefore, show a stronger increase in both mass and radius  due to the magnetic field. This particular feature makes the $M_1$ peak increase in value, leading to a less pronounced minimum between the two peaks of mass ($M_1$ and $M_2$), which characterizes a third family of stars. Depending on the EoS models used and on the intensity of magnetic fields considered, this behavior can eventually result in the elimination of twin stars, as shown in our results for EoS in Set 1.

In addition, we have also studied the case in which strong magnetic fields generate twin-star configurations that otherwise would not exist. This is again related to the increase of mass and radius for the (less compact) hadronic branch. In this particular case, shown for Set 4, the critical mass (that corresponds to a critical central density) becomes a peak of mass $M_1$, which also creates a nearby minimum and, consequently, a third family.

From our results, we can confidently state that the mass minimum generated between the two mass maxima when twin stars are present depends on the central magnetic fields as well as on the compactness of stars in the hadronic and hybrid branches. A more thorough future study considering many twin stars configurations will provide a more model-independent relation between those quantities.

Strong magnetic fields give rise to an instability region on the mass-radius diagram, directly affecting hybrid star configurations by the appearance and/or vanishing of a third family of stars. Together, our conclusions point out the fact that twin stars can only exist as stable objects at specific stages of a magnetar evolution, as either the absence of strong magnetic fields or the presence of very strong ones reduces the number of models/parametrizations that give rise to a mass degeneracy corresponding to stable stars. In the future, we intend to expand our calculations to include different models together with temperature and rotation effects. This will provide more quantitative understandings of how twin stars can be a part of star evolution and how magnetic field decay can generate two families of compact stars.

\begin{acknowledgements}
The authors acknowledge the support from NewCompstar, COST Action MP 1304 and 
HIC for FAIR. 
The authors thank Dr. Mark Alford for fruitful discussions and suggestions. 
V. D. acknowledges support of the National Science Fundation under grant PHY-1748621. 
S.H. is supported by Chandra Award TM8-19002X, the N3AS (Network in Neutrinos, Nuclear Astrophysics, and Symmetries) Research Hub under grant NSF PHY-1630782, and the Heising-Simons Foundation (2017-228 Grant).

\end{acknowledgements}

\bibliography{mag_twins}

\end{document}